\documentclass[12pt, reqno, onecolumn]{amsart}
\usepackage{geometry} 
\usepackage{graphicx}
\usepackage{hyperref}

\title{Reality is both digital and analog}
\author{Alan Forrester\\
112 West Wycombe Road\\
High Wycombe\\
HP12 3AA\\
{\nolinkurl{alan_forrester2@yahoo.co.uk}}}

\begin{document}

\maketitle

\begin{abstract}

I argue that both digital and analog information are important in the foundations of quantum physics. If it is possible for information present in one system to become present in others without being erased in the original system I will say that this information can be copied. I argue that copying is important for understanding issues like causality and that all information that can be copied is digital. I then explain that analog information that cannot be copied can be understood in terms of decision theoretic probability. Finally, I argue that these ideas can help explain the second law of thermodynamics. The arrow of time we experience is the knowledge arrow of time -- the present contains more knowledge, that is, useful or explanatory information, than the past. That knowledge is created by evolutionary processes that involve copying, variation and selection and such processes increase entropy.

\end{abstract}

\section{Introduction}

The laws of physics allow the existence of computers and so allow the existence of systems that can be described in a digital manner. To understand this fact physicists start out with laws of motion that describe how systems evolve over time. When physics is done in this way, the conclusion that the world is analog seems irresistible. All of the laws of motion that have so far been formulated and have survived testing, put continuous quantities front and center. General relativity is a theory of continuous fields in a continuous space-time, as is quantum field theory.

In this view, digital computation is a completely derivative phenomenon. It just so happens that it's possible to arrange for a system to have two stable states. It is also simply a remarkable coincidence that we can put a bunch of these `bits' together and do computations with them, as is the fact that such systems can simulate any other system in quantum mechanics \cite{deutsch89}. I think that this view of the status of computation and so of digital quantities within physics is misleading. There is nothing outside the universe (the whole of physical reality) and so all of the facts about the way the world works have to be instantiated in information that different systems contain about one another. If information \textit{I} is initially instantiated only in system \textit{A} and it then spreads to system \textit{B} while remaining accessible in system \textit{A}, I will say that the information \textit{I} has been \textit{copied}. I shall argue that copying is very important for understanding how physical reality works, that only digital information can be copied according to quantum mechanics, which is the most fundamental theory available to us \footnote{In this essay I will assume that quantum mechanics is universally valid and use the many worlds interpretation of quantum mechanics \cite{everett57}, according to which all of the objects around us exist in multiple versions that are arranged in layers, each of which approximately resembles the universe as described by classical physics.}. Analog information in quantum mechanics is information that can't be copied and dictates how we ought to bet on the outcomes of experiments. I then argue that these ideas can help explain the second law of thermodynamics. The arrow of time we experience is the knowledge arrow of time. Knowledge is created by evolutionary processes that involve copying, variation and selection. Those processes place constraints on the initial and final state of a knowledge creating system that imply that either its entropy grows over time or it increases the entropy of the environment.

\section{Digital and analog information in quantum mechanics}

Let's suppose that I want to explain how a star exploding billions of years ago contributed to the formation of the solar system. This is about as messy, difficult and complicated as any causal explanation can be. The star exploded and this produced a cloud of gas and part of that gas cloud collapsed as a result of gravity,  pressure waves from other exploding stars and so on to form the sun and the solar system. The cloud of atoms that formed the solar system could be traced back in part to that earlier exploding star. So it must be possible to identify that group of atoms over time at least until we reach the point where they were part of that earlier star. Furthermore, it must be possible for that group of atoms to contain some information about the star that produced them that can be copied so that they leave behind information that can help to constrain our theories about the formation of the solar system. If it was impossible in principle to identify that group of atoms as coming from the earlier star, and if they left behind no information that could be copied, it is unclear what it would mean to say that the explosion of that star contributed to the formation of the solar system. So causal explanations involve systems having identity over time and information that can be copied.

In the Heisenberg picture\footnote{I use the Heisenberg picture here because it is more useful for tracking the flow of information between quantum systems \cite{dh00}.}, a quantum system $\mathcal{S}$ is represented by a representative set of operators $O_1, O_2 \dots$ that evolve over time while maintaining the same algebra -- so they evolve unitarily.  That is
\begin{eqnarray}
O_j(t_2) = U^\dagger(t_2, t_1)O_j(t_2)U(t_2, t_1)\nonumber\\
U(t_2, t_1)^\dagger U(t_2, t_1) = \hat{1},
\end{eqnarray}
where $\hat{1}$ is the unit observable and $U(t_2, t_1)$ may be acting on systems other than \textit{S}.

A unitary operator can be written as
\begin{eqnarray}
U = \sum_a \exp(i\phi_a)\hat{P}_a,\nonumber\\
\hat{P}_a\hat{P}_b = \delta_{ab}\hat{P}_a.
\end{eqnarray}
where the $P_a$ form a discrete, i.e. - digital, set of projectors. Let the $X_{ab}$ be operators such that
\begin{eqnarray}
X_{ab}X_{cd} = \delta_{bc}X_{ad},\nonumber\\
X_{aa} = \hat{P}_a,
\end{eqnarray}
i.e. -- they have the same algebra as Dirac dyadics.Then algebra gives
\begin{equation}
U^\dagger X_{cd}U = \sum_{ab}\exp(i(\phi_d-\phi_c))X_{cd},
\end{equation}
and so in general, $U$ only leaves $X_{cd}$ unchanged if $c=d$, i.e. - if $X_{cd}$ is equal to one of the projectors of $U$. If $U$ is degenerate then the $X_{cd}$ for values of $c$ and $d$ for which $U$ is degenerate remain unchanged, but on that particular set of operators $U$ acts as the identity, so nothing changes. So any operator $U$ acts on that remains unchanged. So for an operator that describes a quantum system to remain unchanged when that system interacts non-trivially with other systems it must be a sum of the projectors of the unitary operator acting on it. If that operator only has real eigenvalues then it is an Hermitian operator -- an observable.

Suppose that we have two systems $\mathcal{S}_1$ and $\mathcal{S}_2$ that interact between times $t_1$ and $t_2$. Then if the unitary operator $U$ describing their interaction has the form
\begin{equation}
U(t_1, t_2) = \sum_{a,b} \exp(i\phi_{ab})\hat{P}_{1a}(t_1)\hat{P}_{2b}(t_1),
\end{equation}
where the subscripts 1 and 2 on the operators indicate that they represent descriptors of $\mathcal{S}_1$  and $\mathcal{S}_2$  respectively, then $U$ would leave
\begin{equation}
\hat{A_1}(t) = \sum_a\alpha_a\hat{P}_{1a}(t)
\end{equation}
unchanged. However, it would change $X_{2cd}(t)$
\begin{equation}
X_{2cd}(t_2) = \exp(i(\phi_{ad}-\phi_{ac}))\hat{P}_{1a}(t_1)X_{2cd}(t_1). \label{postmeas}
\end{equation}
And so that interaction would make some of the operators describing $\mathcal{S}_2$  depend on the projectors of $\hat{A}_1(t_1)$. Information about those projectors would be copied from $\mathcal{S}_1$  to $\mathcal{S}_2$ . So only digital information can be copied in quantum mechanics.

Most formulations of quantum mechanics include continuous observables, such as the position observable $\hat{x}$. However, there are no projectors onto individual values of $\hat{x}$ or other continuous observables, but there are projectors onto finite ranges of possible values, see \cite[Sections 3.3, 6.2 and 7.1]{isham95}. So only information about finite ranges of values can be copied and so the positions measured in real experiments are not continuous, they are discrete, i.e. -- they are digital.

However, copying is not the only possibility for transmission of information. Information can be transferred from one system to another with the original system being destroyed, e.g. -- a photon being detected by a CCD. However, information that is going to spread has to be copied. So after the photon has been destroyed some information about it has to be copied if it is going to affect other systems. Copying prevents quantum interference between different values of the copied observable as a result of decoherence. For a review of decoherence see \cite{zurek03}. So there is a distinct version of a system associated with each of the copied projectors that does not interact with the other versions. If between times $t_1$ and $t_2$ a particular version $v$ of a system $\mathcal{S}_1$  copies the information in the projectors $\hat{P}_{2b}(t)$ an observable $\hat{A}_2(t)$ of $\mathcal{S}_2$ all of the information relevant to predicting what will happen is  in $\rho_v\hat{A}(t)$, where $\rho_v$ is a projector that represents the information in $v$ before the interaction. The reason for this is that, from \eqref{postmeas}, the projectors that $\mathcal{S}_1$ had at $t_1$, $\hat{P}_{1a}(t_1)$, one of which is $\rho_v$, change to $\hat{P}_{1a}(t_1)\hat{P}_{2b}(t_1)$, so all of the information about what will happen to $v$ is contained in$\rho_v\hat{A}(t)$. As a result $\rho_v$ only changes when it is updated in the light of new information received by $v$: $\rho_v$ is the relative state of $v$.

This is where analog information starts to become relevant. Suppose that I am in a particular relative state $\rho_v$ and that I am going to measure an observable that has projectors that do not commute with $\rho_v$. That measurement will have multiple outcomes and I will experience each outcome separately, so I somehow have to take all of them into account. In addition, $\rho_v\hat{A}(t)$ does not just contain information about which outcomes will happen. For example, if I am measuring a qubit and $\rho_v = |0\rangle\langle0|$ and $\hat{A}(t) = \tfrac{1}{2}(|0\rangle+|1\rangle)(\langle0|+\langle1|)$, then $\rho_v\hat{A}(t) = \tfrac{1}{2}|0\rangle(\langle0|+\langle1|)$.  And of course we could set up systems to produce any real number from 0 to 1 at the front of $\rho_v\hat{A}(t)$. What does these numbers mean? David Deutsch \cite{deutsch99} has argued that we can use these numbers to make decisions that are rational in a specific decision theoretic sense. When a decision theoretically rational agent in the relative state $\rho_v$ measures an observable $\hat{A}(t)$  whose eigenvalues are the payoff he gets from each outcome that the relevant future version of him will experience he bets as if he is going to receive a payoff:
\begin{equation}
\langle\hat{A}(t)\rangle = tr(\rho_v\hat{A}(t)).
\end{equation}
See also \cite{wallace07} and \cite{forrester07}. This decision theoretic approach to probability does not suffer from the problems of the frequency theory of probability, which states that the probability of an event $x$ under circumstances $y$ is the relative frequency of $x$ over an infinite number of identical experiments conducted in circumstances $y$. But an infinite series of experiments is unphysical, so the frequency theory of probability is problematic in a way that the decision theoretic approach is not.

\section{Thermodynamics and knowledge}

The second law of thermodynamics states that entropy always increases. Entropy can be understood in terms of free energy. Free energy is energy that can be used to do something useful, like the electricity powering my computer. Not all energy is free: the air in my room consists of many molecules that collide with one another, but the energy of those collisions cannot be used to do any useful work unless another system with a different temperature is also used. The free energy $F$ and the entropy $S$ are related by the equation $F= E-TS$,  where $E$ is a term with dimensions of energy, so if $E$ remains constant, $F$ decreases as $S$ increases. 

The second law seems incompatible with the fact that the laws of motion of systems are often time symmetric. There are exceptions, like objects falling into singularities in general relativity, but many interactions are not like this and it is not obvious why the few interactions that are could explain the second law. Many physicists say that the second law is an approximate, statistical law that applies to macroscopic systems, see \cite[Chapter 5 -- 7]{cercignani}. We will never be able to do the calculations required to make atoms undergo anti-entropic motion, which is very improbable for most systems, so the second law will hold with a very high probability, where probability is interpreted in terms of relative frequency. Since the world is quantum mechanical, I will use the decision theoretic approach to probability to avoid the problems of the frequency theory of probability.

However, suppose that our computers improved to the point where we could do the relevant calculations, then could we interfere to undo the second law? This is the concern raised by the Maxwell's demon thought experiment. The experiment involves two boxes of gas with rigid walls, box 1 and box 2, separated by a trapdoor that can slide up frictionlessly. A molecule sized `demon' hides in the box and slides up the partition when a gas molecule is heading into box 1, and so the pressure and temperature of the gas in box 1 increases and that in box 2 decreases. This difference can be used to do useful work and so implies an increase in free energy and a reduction in entropy. This has given rise to a large literature, sampled in \cite{leffrex03}. Most exorcisms of the demon rely on Landauer's principle: the idea that deleting information has a thermodynamic  cost, but Norton \cite{norton05} has argued that these discussions assume the second law applies to the information storage device and so are circular.

The arrow of time we experience is the knowledge arrow of time. The present contains records of the past but the past doesn't contain records of the present and it is this asymmetry that makes the present later than the past. These records contain useful or explanatory information -- knowledge. That knowledge is created by evolutionary processes: processes that involve copying, variation and selection, see \cite[Chapters 1 and 7]{popper79}. Biological evolution creates knowledge about how to copy genes in a particular environmment. Human beings create knowledge about a much wider set of problems, but our knowledge too is created by producing variations on their current knowledge, selecting among those variations according to whether they solve problems and then copying the variations that pass those tests. I shall argue that this process increases entropy of subsystems of the universe.

This approach has some similarities to an argument by Maccone \cite{maccone}, who argued that making records requires the growth of entropy and that the arrow of time is a result of records being created. Jennings and Rudolf \cite{jenningsrudolf} criticised this argument saying that Maccone assumed that classical mutual information was the correct measure of information. To this, I would add that he assumed that the systems involved are initially uncorrelated, i.e. - they are initially in a zero entropy state, so the entropy can only increase. So I don't think this is an adequate explanation of the second law.

Calculations are usually easier in the Schrodinger picture, in which the state changes unitarily, than in the Heisenberg picture, so I am going to switch to the Schrodinger picture for the rest of the essay. In the Schrodinger picture, the state $\rho(t)$ is chosen so that it is equal to the Heisenberg picture state at the time $t=0$ and evolves so that
\begin{equation}
\rho(t_2) = U(t_2, t_1)\rho(t_1)U^\dagger(t_2, t_1).
\end{equation}
In terms of the Schrodinger state, the entropy of a quantum system is 
\begin{equation}
S(\rho(t)) = tr(\rho(t)log\rho(t)). 
\end{equation}
The expectation values of the observables of a subsystem $\mathcal{S}_j$ of a larger system in the state $\rho(t)$ can be calculated from the reduced state of that system
\begin{equation}
\rho_{j}(t) = tr_{not \: \mathcal{S}_j}{\rho(t)}.
\end{equation}
The entropy of a subsystem can be calculated by putting its reduced state into the formula for the von Neumann entropy.

Now suppose that there are two variants of a piece of knowledge that are both involved in a selection process\footnote{The argument can be straightforwardly extended to more than two systems.}. It must be possible for them to be copied independently, otherwise they can't be selected independently. A test itself requires knowledge and the pieces of knowledge to be tested must be subsystems of the system doing the test. Consider a quantum system $\mathcal{S}$ with two subsystems $\mathcal{S}_1$ and $\mathcal{S}_2$: the state of such a system can be written as
\begin{equation}
\rho(t) = \sum_{a,b}p_{ab}|\theta_{ab}\rangle_{1212}\langle\theta_{ab}|,
\end{equation} 
where
\begin{equation}
|\theta_{ab}\rangle_{1212}\langle\theta_{ab}| = \sum_{cdef}\lambda_{abcd}\lambda_{abef}|c\rangle_{11}\langle{e}||d\rangle_{22}\langle{f}|,
\end{equation} 
with $|c\rangle_{11}\langle{e}||g\rangle_{11}\langle{f}| = \delta_{ge}|c\rangle_{11}\langle{f}|$, $|d\rangle_{22}\langle{f}||g\rangle_{22}\langle{h}| = \delta_{fg}|d\rangle_{22}\langle{h}|$ and the $\lambda_{abcd}$ are real numbers such that $\sum_{cd}\lambda^2_{abcd} = 1$. The $\lambda_{abcd}$ can be assumed real due to the Schmidt decomposition. The reduced states of $\mathcal{S}_1$ and $\mathcal{S}_2$ are
\begin{eqnarray}
\rho_1(t) = \sum_{abcde}p_{ab}\lambda_{abcd}\lambda_{abed}|c\rangle_{11}\langle{e}| \nonumber\\
\rho_2(t) = \sum_{abcdf}p_{ab}\lambda_{abcd}\lambda_{abcf}|d\rangle_{22}\langle{f}|.
\end{eqnarray}
If the observable with projectors $|\theta_{ab}\rangle_{1212}\langle\theta_{ab}|$ is measured then the state of $\mathcal{S}$ remains unchanged. But if an observable with the same projectors as $\rho_1(t) $ or $\rho_2(t)$ is measured then the state of $\mathcal{S}$ changes unless $|\theta_{ab}\rangle_{1212}\langle\theta_{ab}|$ is a product of those projectors. So any knowledge bearing system with two knowledge bearing subsystems has a state that can be written as
\begin{equation}
\rho(t) = \sum_{a,b}p_{ab}|a\rangle_{11}\langle{a}||b\rangle_{22}\langle{b}|. \label{knowstate}
\end{equation} 

At the start of the selection process at $t_1$ the state of $\mathcal{S}$ is
\begin{equation}
\rho(t_1) = \sum_{a,b}p_{ab}|a\rangle_{11}\langle{a}||b\rangle_{22}\langle{b}|. \label{start}
\end{equation}
and the reduced states of $S_1$ and $S_2$ are
\begin{eqnarray}
\rho(t_1) = \sum_{a,b}p_{ab}|a\rangle_{11}\langle{a}| \nonumber\\
\rho(t_1) = \sum_{a,b}p_{ab}|b\rangle_{22}\langle{b}|. \label{startred}
\end{eqnarray}
The selection process is a computation involving both $\mathcal{S}_1$ and $\mathcal{S}_2$. This computation must be arranged using current knowledge, which is bound to be imperfect. So even in the case of a classical, i.e. Ð decoherent, computation it will be impossible for epistemological reasons to arrange for the computation to leave $\mathcal{S}_1$ and $\mathcal{S}_2$ in a state of the form \eqref{start}. A slightly more specific argument might help motivate this idea. Knowledge of the orthogonal projectors of the density matrix must be created by a process that produces conjectures about the state and then performing tests of those conjectures. However, those tests do not give perfect access to the expectation values of observables. If the same observables were measured every time any particular version of the tester sees the relative frequencies of different results. The larger the number of trials the more rational it will be (in the decision theoretic sense) to expect that the relative frequencies are close to the probabilities but the relative frequencies will not match the probabilities perfectly, see \cite[Section 4]{forrester07}. The fact that the analog information can't be copied is important for understanding this problem and so for understanding the second law. The experiment is performed with imperfect knowledge from previous rounds of conjecture and criticism and so the testerÕs knowledge of what observable he measures in any particular test will be imperfect and this introduces another source of error. So if the selection process takes place between $t_1$ and $t_2$ the state at $t_2$ is of the form
\begin{equation}
\rho(t_2) = \sum_{a,b}p_{ab}|\theta_{ab}\rangle_{1212}\langle\theta_{ab}|,
\end{equation} 
where
\begin{equation}
|\theta_{ab}\rangle_{1212}\langle\theta_{ab}| = \sum_{cdef}\lambda_{abcd}\lambda_{abef}|c\rangle_{11}\langle{e}||d\rangle_{22}\langle{f}|. \label{end}
\end{equation}
The reduced states of the subsystems at $t_2$ are
\begin{eqnarray}
\rho_1(t_2) = \sum_{abcde}p_{ab}\lambda_{abcd}\lambda_{abed}|c\rangle_{11}\langle{e}| \nonumber\\
\rho_2(t_2) = \sum_{abcdf}p_{ab}\lambda_{abcd}\lambda_{abcf}|d\rangle_{22}\langle{f}|. \label{endred}
\end{eqnarray}
The expectation values of observables of $\mathcal{S}_1$ are now correlated with those of $\mathcal{S}_2$ and they contain information about one another that they did not contain before the interaction and that is what constitutes $t_2$ being later than $t_1$ in this theory. Let
\begin{equation}
|\nu_c\rangle_{11}\langle\nu_c| = \sum_{de}\xi_{cd}\xi_{ce}|d\rangle_{11}\langle{e}|
\end{equation}
where the $\xi_{cd}$ are real c-numbers such that $\sum_{d}\xi_{cd}\xi_{fd} = \delta_{cf}$, then the $|\nu_c\rangle_{11}\langle\nu_c|$ are an orthogonal set of projectors and
\begin{equation}
\sum_c|\nu_c\rangle_{11}\langle\nu_c|\rho_1(t_1) |\nu_c\rangle_{11}\langle\nu_c| = \sum_{abcde}p_{ab}\xi_{cd}\xi_{ce}|d\rangle_{11}\langle{e}|. \label{proj}
\end{equation}
Equation \eqref{proj} is of the same form as equation \eqref{endred}. It can be shown that \cite[Theorem 11.9, p. 515]{NC00}:
\begin{equation}
S(\sum_c|\nu_c\rangle_{11}\langle\nu_c|\rho_1(t_1) |\nu_c\rangle_{11}\langle\nu_c|) \geq S(\rho_1(t_1)).
\end{equation}
So 
\begin{eqnarray}
S(\rho_1(t_2)) \geq S(\rho_1(t_1)),\nonumber\\
S(\rho_2(t_2)) \geq S(\rho_2(t_1)).
\end{eqnarray}

After the computation has been performed the results have to be measured and this too will increase entropy by a similar argument. The only ways to decrease the entropy increase in the knowledge bearing system would be for the interaction to be reversed, which would delete the acquired knowledge, or to export the entropy to another system. But this would just mean that the entropy of some other system would increase and so it would not allow the evasion of the second law. So the growth of knowledge is correlated with the growth of entropy.

\section{Conclusion}

Both analog and digital information are important in physics. All information that can be copied is digital, but this information does not capture all of the information in a system and the rest is characterised by analog information, which acts as decision theoretic probability. Together they help to explain the second law of thermodynamics. The digital information that can be copied contains knowledge that explains the arrow of time. The analog information that cannot be copied contributes to making knowledge imperfect, and this contributes to the growth of entropy.

\textit{Acknowledgements} -- I would like to thank David Deutsch for comments on an earlier draft of this essay.

\end{document}